# Observation of zero coefficient of friction above a critical pressure

Weipeng Chen[1,2], Tielin Wu[1,2], Yelingyi Wang[1,2], Deli Peng[3], Jin Wang[4], Zhanghui Wu*[1,2], Quanshui Zheng*[1,2,3,5]

1 Center for Nano and Micro Mechanics, Tsinghua University, Beijing 100084, China;

2 Department of Engineering Mechanics, School of Aerospace Engineering, Tsinghua University, Beijing 100084, China;

3 Institute of Superlubricity Technology, Research Institute of Tsinghua University in Shenzhen, Shenzhen 518057, China;

4 International School for Advanced Studies (SISSA), Via Bonomea 265, 34136 Trieste, Italy;

5 Tsinghua Shenzhen International Graduate School, Shenzhen 518057, China.

*Corresponding author: Zhanghui Wu and Quanshui Zheng

Email: wuzh1995@mail.tsinghua.edu.cn; zhengqs@tsinghua.edu.cn

## Abstract

Self-superlubricity is a highly anticipated phenomenon where certain solid pairs in contact, without lubricant, exhibit zero wear and virtually null static friction and coefficient of friction (CoF). We present the first experimental observation of self-superlubricity in a microscale single-crystalline graphite flake in contact with a nanoscale-rough Au substrate, achieved when the applied normal pressure exceeds a critical threshold. Theoretical analysis revealed that substrate roughness impedes full contact at low pressures, but increasing the pressure induces a transition to full contact, enabling self-superlubricity. We established a dimensionless criterion for this critical pressure, further validated by observing self-superlubricity between graphite and an atomically smooth sapphire substrate without requiring additional pressure. This breakthrough introduces a transformative principle for next-generation microsystems such as micro/nanoscale generators, motors, oscillators, sensors, etc., enabling reduced power consumption and extended operational lifetimes in applications such as 6G communication, humanoid robotics, and unmanned aerial vehicles.

## Introduction

Since Feynman's seminal 1959 lecture[1], "*Plenty of Room at the Bottom*", the transformative potential of microscale technologies has been widely acknowledged. Building on this vision,

microsystems—integrating sensing, communication, processing, actuation, and micro-energy components—have emerged as "the inevitable, logical development of microelectronics into non-electronic areas"[2]. Despite this progress, traditional semiconductor and MEMS-based microsystems face significant technological bottlenecks and fundamental challenges, particularly in addressing the increasing demands for low power consumption[3] and extended operational lifetimes in next-generation industries such as 6G communication, humanoid robotics, and unmanned aerial vehicles[4,5]. Over the past two decades, the emergence of superlubricity[6,7,8,9,10,11,12,13,14,15]—characterized by extremely low friction and wear—has offered a promising solution to these challenges and introduced a much-needed novel technological principle for microsystems[16,17]. This advancement enables a new degree of freedom for interface sliding at micro- and nanoscales[18,19,20], fundamentally redefining microsystem design principles and pioneering a new frontier: superlubric microsystems, miniaturized systems incorporating superlubricity components[21,22].

Historically, the concept of superlubricity stems from "lubricity"[23], with most implementations relying on solid or liquid lubricants, which often permit wear. While, when scaled down to the micro- and nanoscale, most liquid lubricants become excessively viscous or fail to withstand the required pressure[13,24], leading to failure. Besides, even minimal wear can be catastrophic for high-frequency micro-devices, where precision and durability are paramount[25]. This raises a critical question: can the "self-superlubric" state—defined by zero wear and virtually null static friction and coefficient of friction (CoF)—be achieved solely through intrinsic contact?

Recent advancements suggest this may be possible.

(i) **Zero wear**: Experimental demonstrations of wear-free sliding over 100 kilometers between graphite and diamond-like carbon (DLC) substrates[26], and the stability of superlubricity under pressures up to 3.74 GPa for graphite/tungsten interfaces, 9.45 GPa for graphite/graphite interfaces[27], highlight the possible wear-free nature of intrinsic superlubric interfaces.

(ii) **Null static friction**: The elimination of edge pinning by selectively etching amorphous carbon along graphite edges has resulted in the disappearance of static friction force[28,29], suggesting that static friction is not an intrinsic property of interfacial friction.

(iii) **Null CoF**: Exceptionally low or even negative CoFs have been reported in two-dimensional (2D)/2D systems, including values such as $1.2 \times 10^{-5}$, $6 \times 10^{-6}$, and $-7 \times 10^{-5}$ for graphite/graphite interfaces[30,31,32], $1.4 \times 10^{-4}$, $-7 \times 10^{-4}$ for graphite/hBN interfaces[33,34], and $5 \times 10^{-5}$ for $MoS_2$/$MoS_2$ interfaces[35]. But previous studies have largely overlooked the physical significance of such extreme CoFs, focusing instead on achieving the engineering threshold for superlubricity (< 0.01).

In this work, we report the first observation of the long-anticipated superlubric state that simultaneously exhibits all three defining characteristics and introduce the term self-superlubricity to describe it. This state is achieved through direct contact between two solid surfaces without lubricants and is distinguished by zero wear and virtually null static friction and CoF when the normal pressure exceeds a critical threshold. Interestingly, the behavior of the CoF in this state reveals a remarkable parallel to the critical threshold phenomena commonly observed in superconductivity[36].

## Results

**Experimental Observation**

We investigated the frictional properties of intrinsic 2D/three-dimensional (3D) interfaces using arrays of 4 μm square graphite mesas (Supplementary Fig. S1) and 2.5 μm diameter circular Au pillars (Supplementary Fig. S2) to construct pure in-plane contacted graphite/Au interfaces (Methods). The preparation methods are detailed in the Methods section. As shown in Fig.1a, self-retracted graphite flakes were transferred onto the Au pillars[32], assembling the 2D/3D SSL heterojunction under study. In the aforementioned constructed interface, graphite provides an atomically smooth, single-crystalline van-der-Waals (vdW) surface (see Fig.1b), while the Au pillar offers a nanoscale rough surface morphology (see Fig. 1c, roughness, $R_q \sim 1$ μm), detailed in Methods. Unlike traditional superlubric junctions[26,33], the edges of the graphite flakes are suspended, as illustrated in Fig.1d, thereby eliminating the substantial influence of the amorphous carbon regions at the flake edges (approximately 30 nm wide) on the interface's frictional properties[28] (Supplementary Fig. S3). Subsequently, this junction was transferred to the AFM system for friction force testing (Methods), where the graphite flake was pressed by the AFM tip and slid reciprocally over the surface of the Au pillar. The frictional forces $F_r$ at the graphite/Au interface were inferred from the measured friction loops as depicted in Fig.1e, with the displacement set to 100 nm to avoid edge contact. Additionally, a comparable system using sapphire pillars was fabricated to provide a 3D substrate surface with atomically smooth ($R_q \sim 0.3$ μm) for comparative experiments and analysis (Supplementary Fig. S4).

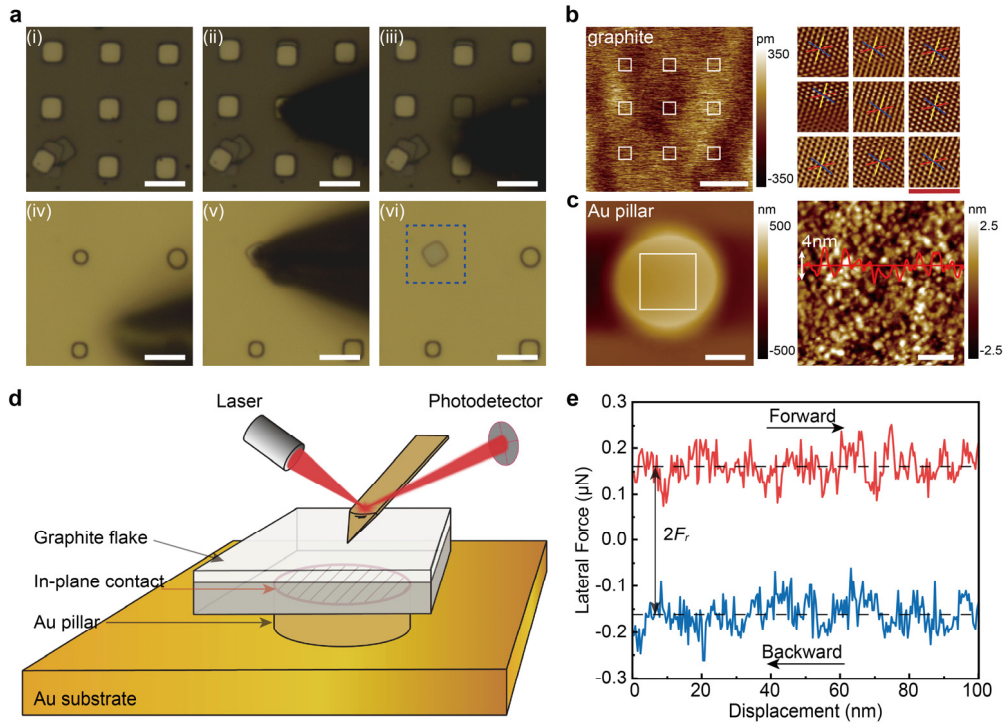

**Fig. 1: Experimental set-up. a,** Optical micrographs showing the process to obtain the graphite/Au heterojunction. Scalebar, 8 μm. **b,** The topography (left panel) and corresponding lattice orientation (right panel) of the lower surface at the self-retracted graphite interface. Scalebars, 1 μm and 2 nm respectively. **c,** The topographies of Au pillar (left panel) and its nanoscale rough surface (right panel). Scalebars, 1 μm and 300 nm respectively. **d,** Schematic diagram

of the experimental setup for measuring the interfacial friction between graphite flake and Au pillars, with pure in-plane contact. **e,** A typical friction loop used to extract the friction force of relative sliding at the graphite/Au interface, with a displacement set to 100 nm.

Through this edge-suspended structural design, we observed distinct friction properties beyond previously reported graphite-based superlubricity, at both graphite/Au and graphite/sapphire interfaces, as shown in Figs. 2a and 2b. Two remarkable features emerged:

(1) For the first time, pressure-independent friction was observed on 3D substrates with a certain roughness. Both the latter half of the Au substrate experiments (Fig. 2a) and the entire span of sapphire substrate experiments (Fig. 2b) exhibited near-zero CoFs, typically on the order of $\pm 0.0001$ or less. The occurrence of extremely low or even negative CoFs is attributed to experimental uncertainties, which we estimate to be up to 0.001 (see Supplementary Discussion 1). Previously, similar indications had been only observed in graphite-based superlubric systems on 2D substrates (e.g., graphite[30, 31, 32, 37], hBN[33, 34], and $MoS_2$[38, 39], Fig. 2c), while edge contacts on 3D substrates[21, 26, 38, 40] consistently exhibited CoFs on the order of 0.001 (Fig. 2d). The fundamental nature behind these phenomena remained unexplored. Although both CoF ranges were traditionally considered within the superlubricity regime, they represent two distinct characteristics: CoFs around 0.001 show a clear positive correlation between frictional force and normal pressure, while CoFs as low as ±0.0001 or even lower indicate undetectable pressure dependence. By eliminating edge amorphous carbon, graphite-based systems achieved null CoFs on both 2D and 3D substrates. This discovery challenges the traditional CoF threshold for superlubricity, prompting an exploration of the underlying mechanisms.

(2) An unexpected nonlinear transition in frictional force with increasing normal pressure was observed on 3D substrates with nanoscale roughness. As shown in Fig. 2a, the pressure dependency of frictional force at graphite/Au interfaces can be segmented into two linear regions, based on the pressure intervals, with CoF being from 0.00230 to -0.00001, and from 0.00422 to 0.00034. Additional data points from multiple samples are provided in Supplementary Fig. S5, and the summary of CoF transitions is consolidated in Fig. 2d. This analysis reveals that as pressure increases, the graphite/Au interface transitions from a state of positive pressure-dependence to pressure independence. This transition offers crucial insights into the fundamental origin of vanishing CoFs in intrinsic superlubric interfaces. Further exploration of this phenomenon is presented in the following theoretical model section, where we demonstrate that the null CoF arises from the intrinsic full contact achieved at the superlubric interface.

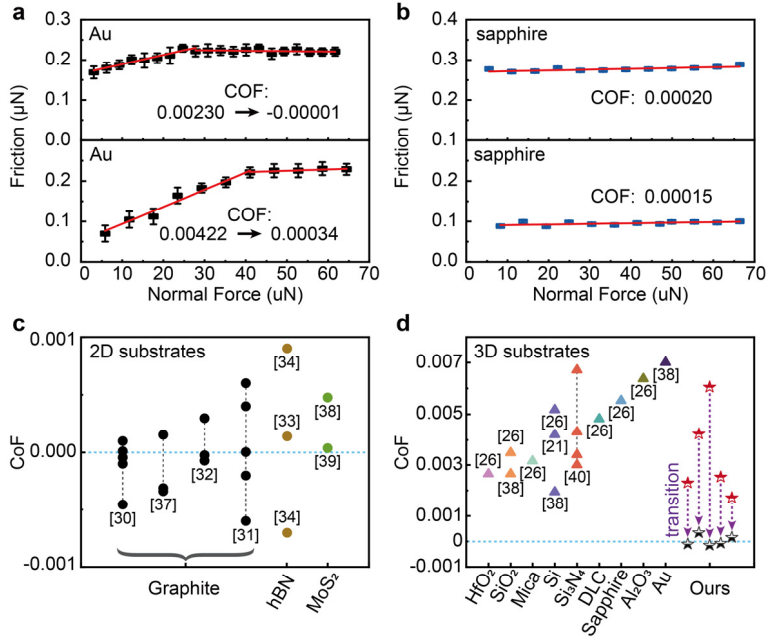

**Fig. 2: Discovery of pressure-independent friction and nonlinear transition in CoF on 3D Substrates. a,** CoF measurements on the graphite/Au interface, revealing two distinct linear regions of pressure dependency. **b,** Pressure-independent friction observed on the graphite/Sapphire interface, with CoF consistently on the order of 0.0001. **c,** Comparison with frictional behavior on 2D substrates, where similar negligible CoF (±0.0001) are observed. **d,** Summary of frictional characteristics on 3D substrates: systems with edge contacts exhibit CoF consistently greater than 0.001, showing clear pressure dependence (marked by triangles); whereas systems with non-edge-supported Au pillars exhibit a transition from CoF values around 0.001 to values as low as ±0.0001 as pressure increases (marked by pentagrams).

Furthermore, we verified the remaining characteristics of self-superlubricity—virtually null static friction and zero wear—through long-cycle sliding friction experiments. First, as shown in Fig. 3a, the friction loop does not display any discernible static friction peaks, indicating the vanishing static friction force of the system. Second, the friction force measured over 2000 sliding cycles, as illustrated in Fig. 3b, remains remarkably stable at approximately 0.28 μN, further supporting the robustness of this superlubric state. Following the sliding experiments, we conducted in situ topographical characterization of the Au surface. As depicted in Fig. 3c and 3d, no wear debris or morphological damage was observed, with every microstructure maintaining one-to-one correspondence. Lastly, we flipped the bottom surface of the graphite flake and performed Raman spectroscopy to assess structural defects (Supplementary Fig. S6). As shown in Fig. 3e, no defect-related D peak (1350 cm$^{-1}$) were detected in the Raman spectra at any of the nine selected points across the contact region (Methods).

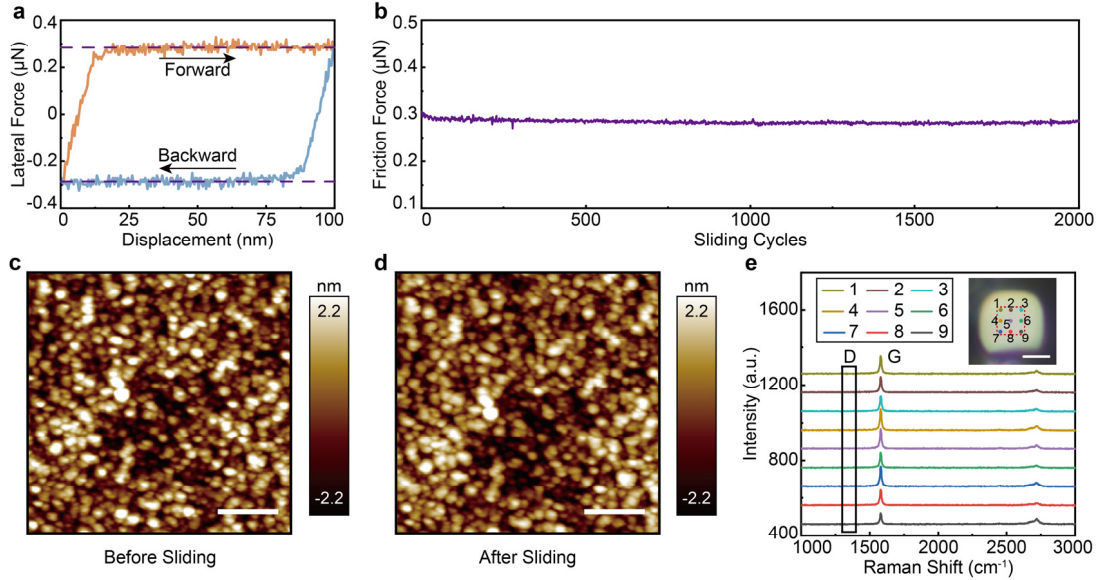

**Fig. 3: Characterization of virtually null Static Friction and Zero Wear. a,** The friction loop measured under a normal load of 73 μN, with a distance of 100 nm and a frequency of 1 Hz, exhibits no discernible static friction peak. **b,** The friction force recorded over 2000 consecutive sliding cycles remains remarkably stable, further confirming the robustness of this state under extended sliding conditions. **c,d,** Topographical images of the Au surface taken before (**c**) and after (**d**) the 2000 sliding cycles demonstrate complete preservation of local surface features, with every microstructure maintaining one-to-one correspondence. Scalebars, 300 nm. **e,** Raman spectroscopy of the bottom surface of the graphite flake after 2000 sliding cycles reveals no detectable structural degradation. Inset, scalebar, 2 μm.

**Theoretical Model**

To elucidate the underlying mechanism, we considered the pressure-dependent evolution of the contact state at the graphite/Au interface using contact mechanics theory[41] (detailed in Supplementary Discussion 2). The interface model (Fig. 4a) features a rough Au surface represented by sinusoidal modulation against a smooth graphite flake underlayer. Given the low normal stiffness of graphite, external pressure $P_{zz}$ will effectively cause the interfacial deformation, resulting in a better conformity to the Au substrate[26]. Here, we introduce the concept of the contact ratio $\chi$ defined as the ratio of the contact width $a$ to the period width $L$ of a single sinusoidal cycle ($\chi = a/L$, with $\chi = 1$ representing full contact). The pressure $P_{zz}$ required to achieve a given $\chi$ depends on material properties, thickness $H$, peak-to-valley value of the surface roughness $h$, and adhesion energy $\gamma$, as described by: (Noting that the ratio $\gamma/E$ is a length unit.)

$$P_{zz} = \frac{Eh}{H}\zeta(\chi) - \frac{2\gamma}{h}\eta(\chi). \qquad (1)$$

Here $E$ is the effective Young's modulus of the graphite flake/Au pillar interface, and $\zeta(\chi)$, $\eta(\chi)$ are dimensionless functions related to the contact ratio $\chi$, as illustrated in Fig. 4b inset (their specific expressions and derivation process are detailed in Supplementary Discussion 2). According to Eq. (1), aside from the intrinsic material properties ($E$ and $\gamma$), the contact state of the interface will heavily depend on $H$ and $h$. For the case $h/H\sim0$ (either the flatness is super tiny or the bulk thickness is super large), full contact can be easily achieved. Otherwise (either $H$ is small or $h$ is large), the full contact will depend very much on the adhesion $\gamma$.

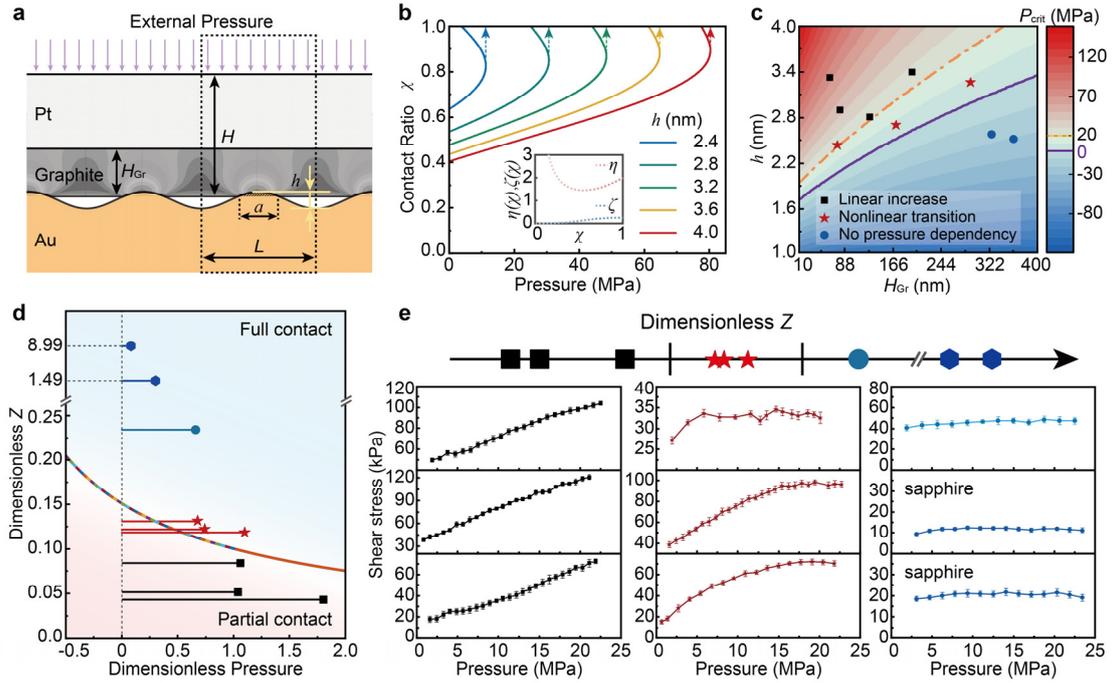

**Fig. 4: Pressure-driven evolution of contact state and implications for frictional behavior. a,** Schematic of the graphite/Au interface, with the dashed box showing a single contact unit. **b,** Interface contact ratio $\chi$ as a function of pressure $P_{zz}$, derived from Eq. (1). Curves represent varying peak-to-valley surface roughness values $h$ (as marked in (**a**)). Inset: Relationships between the dimensional parameter $\zeta$, $\eta$ and the contact ratio $\chi$. **c,** Phase diagram of critical pressure $P_{crit}$ as a function of $h$ and graphite thickness $H_{Gr}$. Three frictional behaviors are identified: linear friction (black diamonds), nonlinear transitions (red pentagrams), and pressure-independent friction (blue circles). **d,** Contact state phase diagram using dimensionless parameters $Z$ (interface configuration) and $p_{zz}$ (applied pressure). The dimensionless criterion denoted by the colored solid line, divides the phase diagram into full contact (filled in blue) and partial contact (filled in red) states. Horizontal lines indicate experimental data. **e,** Experimental data validating model predictions and friction state transitions in (**d**).

By incorporating our experimental parameters (Supplementary Table S1) into Eq. (1), we modeled the variation of the interface contact ratio $\chi$ during compression, revealing a non-monotonic relationship with pressure (Fig. 4b). Notably, interfacial adhesion energy induces an initial contact area even in the absence of external pressure, consistent with prior reports. This explains why the linear relationship between friction force and pressure deviates from the origin. As pressure increases, the contact area expands until a critical pressure threshold, $P_{crit}$, is reached. Beyond this point, the interface transitions abruptly to full contact, as marked by the dashed arrow in Fig. 4b. This non-monotonic behavior arises from the interplay between deformation energy, which resists contact, and adhesion energy, which drives the transition. Eq. (1) encapsulates this balance, highlighting the competing contributions of these energies during the compressive process.

To account for variations in experimental parameters across samples, we developed a phase diagram (Fig. 4c) showing the critical pressure $P_{crit}$, as a function of the graphite layer thickness $H_{Gr}$ and the average peak height $h$ of the Au surface roughness. The critical pressure $P_{crit}$ signifies the minimum pressure required to achieve full contact. A key feature of this phase diagram is the boundary where $P_{crit} = 0$ (solid purple line), which separates two distinct regimes: in Region 1 ($P_{\text{crit}} < 0$), full contact occurs spontaneously without external pressure, resulting in pressure-independent friction; whereas in the top-left region ($P_{\text{crit}} > 0$), external pressure is required to achieve full contact. The top-left region is further divided based on the maximum experimental pressure (~20 MPa, marked by the orange dashed line in Fig. 4c): in Region 2 ($0 < P_{\text{crit}} < 20$ MPa), full contact can be achieved within the experimental range, enabling observations of a nonlinear friction transition with increasing normal pressure. In contrast, in Region 3 ($P_{\text{crit}} > 20$ MPa) full contact cannot be reached, leading to persistent partial contact and a monotonic increase in friction with pressure.

To validate the correlation between observed friction behaviors and predicted contact states, we mapped additional experimental results onto the phase diagram of Fig. 4c. We conducted systematic friction-pressure measurements while characterizing Au surface roughness and graphite thickness (Supplementary Fig. S7). The results categorized into three behaviors: linear positive correlation (black diamonds), nonlinear transition (red pentagrams), and weak correlation (blue circles). These data points align well with their theoretically predicted regions, confirming that the unique frictional characteristics of these intrinsic superlubric interfaces originate from contact state transitions.

To generalize our findings, we performed a dimensionless analysis (Supplementary Discussion 5), and derived a universal criterion for 2D/3D interface contact states (Fig. 4d, colored solid lines). The dimensionless critical pressure $p_{\text{crit}}$ depends solely on the dimensionless interface configuration parameter $Z$, which relates to surface roughness and layer thickness. This criterion enables contact state determination based on interface configuration ($Z = 2\gamma H/Eh^2$) and applied dimensionless pressure ($p_{zz} = hP_{zz}/2\gamma$). In Fig. 4d, the light blue region represents full contact ($p_{zz} > p_{\text{crit}}$), while the light red region indicates partial contact. The dimensionless analysis not only consolidates interface configuration into parameter $Z$ but also captures real-time pressure variations during loading. The correlation between experimental friction characteristics and predicted contact states reveals consistent patterns: linearly increasing friction data (black squares, Au) remains within the partial contact region, while nonlinearly transitioning data (red pentagrams, Au) shows evolution from partial to full contact. Pressure-independent friction data (blue circles, Au; dark blue hexagons, sapphire) consistently falls within the full contact region (Fig. 4d and 4e).

## Discussions

In this work, we clearly define the concept of self-superlubricity and report its first experimental observation in microscale single-crystalline graphite flakes sliding against Au or sapphire substrates with a certain roughness. Building on insights from previous studies, we hypothesize that such an extreme state of self-superlubricity could similarly be observed at vdW interfaces in 2D/2D or 2D/3D systems under suitable conditions. The essence of this "zero-friction" state lies in the intrinsic vdW full contact formed between two atomically flat surfaces, governed by the following principles: 1) During lateral shear motion between sufficiently in-plane rigid surfaces, the vdW interactions exhibit mutual cancellation systematically, ensuring ultralow frictional dissipation. 2) The absence of strong interfacial bonds (e.g., metallic, covalent, or even hydrogen bonds) guarantees zero wear and virtually null static friction force, as the interactions are limited to vdW forces. 3) The stable full-contact state directly leads to the pressure-independent null CoF and other robust mechanical properties.

This mechanism of intrinsic vdW full contact fundamentally differentiates self-superlubricity from conventional lubrication strategies, establishing a groundbreaking lubricant-free mode of operation. To encapsulate the intrinsic, self-contacting nature of this phenomenon, we propose the introduction of the prefix "self" to superlubricity, defining it as self-superlubricity. This terminological refinement not only reflects our deeper understanding of the mechanism but also paves the way for reimagining and advancing the design of superlubric systems in the future. Moreover, self-superlubricity associated technologies represent the optimal pathway toward realizing superlubric microsystems, having the potential to emerge as a core enabling technology for the fourth industrial revolution, catalyzing transformative advances across multiple technological domains.

# Methods

**Formation of graphite flake/Au pillar heterojunctions.** The fabrication of graphite/Au frictional contact (2D/3D junction) followed the protocol depicted in Fig. 1a (i-vi): (1) Cleave the graphite flake with self-retracted motion from the original graphite mesa of 4 μm by tungsten probe. (2) Transfer the graphite flake above the Au pillar with the diameter of 2.5 μm. (3) Assemble the graphite/Au SSL junction with only in-plane contact, based on their interfacial adhesion.

**Surface Characterization.** 1) The morphology of the graphite surface was captured in contact mode using an atomic force microscope (AFM, Cypher S-Oxford Instruments) (0.11 N/m, Au-coated tips, NT-MDT & TipsNano). Subsequently, the system was switched to lateral force mode to scan nine selected surface points. Frictional data was Fourier-transformed from the recorded stick-slip traces to generate a crystal phase map of the interface. 2) The morphology of the Au pillar was analyzed in tapping mode using an AFM (Cypher S-Oxford Instruments) (22~100 N/m, Au-coated tips, NSG30). 3) To evaluate possible damage to the graphite flake surface, Raman spectroscopy was conducted. The graphite flake was lifted using AB adhesive to overcome vdW adhesion with the Au substrate. The flake was then flipped 180° for Raman characterization using a LabRAM HR Evolution spectrometer (HORIBA). The setup included a 532 nm laser wavelength, 1800 gr/mm grating, 0.1 cm$^{-1}$ spectral resolution, 4 s acquisition time, and 1 μm spot diameter.

**Friction measurement.** The friction experiments were performed using a commercial NTEGRA upright atomic force microscope (AFM) from NT-MDT, operating in ambient conditions. The AFM setup incorporated a 100μm XYZ piezoelectric displacement platform, a high numerical aperture objective lens (×100, Mitutoyu), and a visualized AFM tip (VIT\_P/IR, TipsNano). We selected an AFM probe with a force constant ranging from 25 to 95 N/m and a resonance frequency between 200 and 400 kHz. To measure friction at SSL interface, we applied pressure to the graphite flake using the AFM probe, inducing relative sliding between the graphite and the Au substrate. The AFM tip's sensitivity was calibrated in situ: normal sensitivity was determined using the Sader method, while lateral sensitivity was calibrated using a diamagnetic levitation spring system (Supplementary Discussion 6).